\RequirePackage{fix-cm}
\documentclass[10pt,journal,compsoc]{IEEEtran}

\usepackage{rotating}
\usepackage{subcaption}

\usepackage{cite}
\usepackage{graphicx}
\usepackage{multirow}
\usepackage{hhline}
\usepackage{comment}
\usepackage[dvipsnames]{xcolor}
\usepackage[colorlinks=false, linkcolor=blue, citecolor=blue, bookmarks=true,pagebackref=false]{hyperref}
\usepackage{hyphenat}
\usepackage{subcaption}
\usepackage{float}
\usepackage{caption}
\usepackage{fancybox}
\usepackage{array,booktabs}
\usepackage{marvosym} \usepackage{balance}
\usepackage{ragged2e}
\usepackage{makecell}
\usepackage{bigstrut}
\newcolumntype{P}[1]{>{\centering\arraybackslash}p{#1}}
\usepackage{amssymb,mathtools}
\newcommand{\rqbox}[1]{
	\begin{center}
	\vspace{0cm}
	\cornersize{.2}
	\setlength{\fboxsep}{5pt}
	\ovalbox{\begin{minipage}{3.3in}
	{\em #1}
	\end{minipage}}
	\vspace{-0.1cm}

	\end{center}
}

\newcommand{\code}[1]{{\texttt{#1}}}

\begin{document}

\title{Bounties in Open Source Development on GitHub: A Case Study of Bountysource Bounties}

\author{Jiayuan~Zhou,
        Shaowei~Wang,
        Cor-Paul~Bezemer,
        Ying~Zou,~\IEEEmembership{Member,~IEEE,}
        and~Ahmed~E.~Hassan,~\IEEEmembership{Member,~IEEE}
\IEEEcompsocitemizethanks{\IEEEcompsocthanksitem J. Zhou, S. Wang and A. E. Hassan are with the Software Analysis and Intelligence Lab (SAIL), Queen's University, Kingston, Ontario, Canada.\protect\\
E-mail: {jzhou,shaowei,ahmed}@cs.queensu.ca
\IEEEcompsocthanksitem Cor-Paul Bezemer is with the Department of Electrical and Computer Engineering, University of Alberta, Edmonton,
              AB, Canada
E-mail: bezemer@ualberta.ca
\IEEEcompsocthanksitem Y. Zou is with the Department of Electrical and Computer Engineering, Queen's University, Kingston, Ontario, Canada.\protect\\
E-mail: ying.zou@queensu.ca
\IEEEcompsocthanksitem Shaowei Wang is the corresponding author.
}

}

\IEEEtitleabstractindextext{

\begin{abstract}
Due to the voluntary nature of open source software, it can be hard to find a developer to work on a particular task. For example, some issue reports may be too cumbersome and unexciting for someone to volunteer to do them, yet these issue reports may be of high priority to the success of a project. To provide an incentive for implementing such issue reports, one can propose a monetary reward, i.e., a bounty, to the developer who completes that particular task.
In this paper, we study bounties in open source projects on GitHub to better understand how bounties can be leveraged to evolve such projects in terms of addressing issue reports.

We investigated 5,445 bounties for GitHub projects. These bounties were proposed through the Bountysource platform with a total bounty value of \$406,425.
We find that 1) in general, the timing of proposing bounties and the bounty-usage frequency are the most important factors that impact the likelihood of an issue being addressed. More specifically, issue reports are more likely to be addressed if they are for projects in which bounties are used more frequently and if they are proposed earlier. 2) The bounty value that an issue report has is the most important factor that impacts the issue-addressing likelihood in the projects in which no bounties were used before. Backers in such projects proposed higher bounty values to get issues addressed. 3) There is a risk of wasting money for backers who invest money on long-standing issue reports.
\end{abstract}

\begin{IEEEkeywords}
Bountysource, Bounty, GitHub, Issue Report, Open Source Software
\end{IEEEkeywords}

}

\maketitle

\section{Introduction}\label{sec:introduction}

Software projects often use issue tracking systems to store and manage issue reports. Developers or users submit issue reports to report bugs or request new features, and wait for these issues to be addressed. However, some issue reports may never be addressed.
For example, developers may avoid addressing issues that they consider too low priority, or difficult to implement. To encourage developers to address such issue reports, a \textit{bounty} can be proposed by one or more \textit{backers} for the issue reports.

A bounty is a monetary reward that is often used in the area of software vulnerabilities. Prior studies examined the impact of bounties on vulnerability discovery \cite{zhao2017devising, hata2017understanding,finifter2013empirical}. Finifter et al.~\cite{finifter2013empirical} suggested that using bounties as an incentive to motivate developers to find security flaws is more cost-effective than hiring full-time security researchers.

Bounties are now being used to motivate developers to address issue reports, e.g., to fix bugs, to improve performance, or to add new features.
$Bountysource$\footnote{\url{https://www.bountysource.com}} is a platform for proposing bounties for open source projects across multiple platforms (e.g., GitHub) which currently has more than 46,000\footnote{\url{http://bit.ly/2RkDeCc}} registered developers. Although bounties are used in the issue-addressing process, the relationship between bounties and this process is not yet understood. For example, it is unclear whether a bounty improves the likelihood of an issue being addressed (i.e., the issue-addressing likelihood) in projects. By understanding this relationship, we could provide insights on how to better leverage bounties to evolve open source projects, and on how to improve the usability and effectivity of bounty platforms.

In this paper, we study 3,509 issue reports with 5,445 bounties that were proposed on Bountysource from 1,203 GitHub projects, with a total bounty value of \$406,425. We first examine the impact of the frequency of bounties (i.e., the bounty-usage frequency) being used in projects and the timing of proposing bounties on the likelihood of an issue being addressed. We found that:
\begin{enumerate}
    \item Bounty issue reports are more likely to be addressed in projects which are using bounties more frequently. Backers of the projects in which bounties were not used before proposed higher bounty values to get issues addressed.
    \item Bounty issue reports for which bounties were proposed earlier are more likely to be addressed. Additionally, there is a risk of wasting money for backers who invest money on long-standing issue reports.
\end{enumerate}

To understand if there are other factors that have an impact on the issue-addressing likelihood, we use logistic regression to study the relationship between 27 factors (including the timing of proposing a bounty and the bounty-usage frequency of a project) along 4 dimensions (i.e., project, issue, bounty, and backer) and the issue-addressing likelihood. We found that for bounty issue reports:
\begin{enumerate}
\item The timing of proposing bounties and the bounty-usage frequency are the two most important factors that impact the issue-addressing likelihood.
\item The total bounty value that an issue report has is the most important factor that impacts the issue-addressing likelihood in the first-timer projects.
\end{enumerate}

We also performed a manual study on the addressed bounty issue reports in which the bounty remained unclaimed (i.e., the cases in which bounties were ignored by developers). We found that some developers addressed an issue cooperatively, making it difficult to choose a single developer that would be awarded the bounty. In addition, some developers are not driven by money to address issues.

Based on our findings, we have several suggestions for backers and the Bountysource platform. For example, backers should be cautious when proposing small (i.e., $< \$100$) bounties on long-standing issue reports since the risk of losing the bounty exists. Bounty platforms should consider allowing for splittable multi-hunter bounties.

The rest of the paper is organized as follows.
Section~\ref{background} presents background information about GitHub and Bountysource.
Section~\ref{dataset} describes our data collection process.
Section~\ref{prestudy} introduces our preliminary study.
Section~\ref{rq1} and Section~\ref{rq2} study the impact of the bounty on the issue-addressing likelihood in terms of projects' bounty-usage frequency and the timing of proposing bounties.
Section~\ref{rq3} investigates more factors that may potentially affect the issue-addressing likelihood.
 Section~\ref{sec:dis} studies the closed-unpaid bounty issue reports and discusses the implications of our study. Section~\ref{sec:threats} discusses the threats to validity of our study. Section~\ref{relatedwork} introduces related work. Finally, Section~\ref{conclusion} concludes our study.

\section{Background}
\label{background}
In this section, we briefly introduce the issue tracking system on GitHub and the open source bounty platform Bountysource.

\subsection{Issue tracking system on GitHub}

The issue tracking system (i.e., ITS) on GitHub helps developers to manage their project with issues. Users and developers can report bugs or request new features by posting an issue report on the issue tracking system. There are two statuses of an issue report: ``open'' and ``closed''. ``Open'' indicates that the issue report is still active and is waiting to be addressed. ``Closed'' indicates that the issue report has been closed. The most common reason for closing an issue report is that the issue has been addressed, but it could also have other reasons (e.g., duplicated issue reports).
Users can attach free-text labels to issue reports to indicate the category of an issue report.
A issue report contains a title to summarize the issue and a description that describes the issue in detail.
Developers can discuss an issue report by leaving comments, which can include code snippets, links or images to improve the description.

\subsection{Bountysource}
Bountysource is a platform on which users can pledge a monetary incentive (a \textit{bounty}) to address an issue report of an open source project.
There exist two roles on Bountysource: the bounty backer and the bounty hunter roles.

\noindent\textbf{Bounty backers}, which may be anonymous, are users or developers who propose bounties for issue reports. The backer can set an expiration period for bounties that are over \$100. When the bounty expires, the money is refunded to the backer; otherwise, the bounty stays with the issue report until someone claims it. Bounties that are smaller than \$100 are not refunded if they remain unclaimed. One issue report can have multiple bounties from one or more backers.

\noindent\textbf{Bounty hunters} are developers who address issue reports that have bounties. Once a developer claims to have addressed an issue report, its bounty backer(s) can choose to accept (no response will be taken as an acceptance) or reject the claim. In this situation, backers have two weeks to make the decision (accept or reject). If no backer explicitly rejects the claim, the bounties will be paid to the developer automatically.
Multiple bounty hunters can work on an issue report at the same time, but the bounties of an issue can only be rewarded to one bounty hunter.

\begin{figure}[t]
   \centering\includegraphics{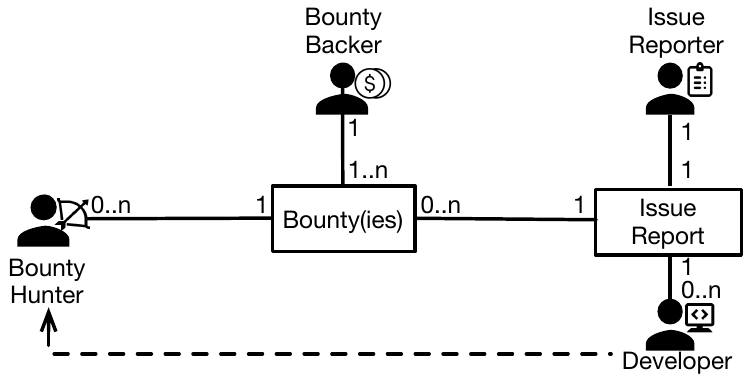}
      \vspace{-0.1in}
  \caption{The relationships between entities that are involved in the  bounty process.}
  \label{RelationEntity}
  \vspace{-0.1in}

\end{figure}

Figure~\ref{RelationEntity} summarizes the relationships between the bounty, the bounty backer, the bounty hunter, the issue report, the issue reporter, and the developer. Basically, when an issue report is submitted by an issue reporter, one or more bounty backers can propose bounty(ies) on the issue report. One or more developers of the issue report can choose to become bounty hunters to address the issue report but only one bounty hunter can get the bounty(ies).

\begin{figure}[t]
   \centering\includegraphics[width=9cm]{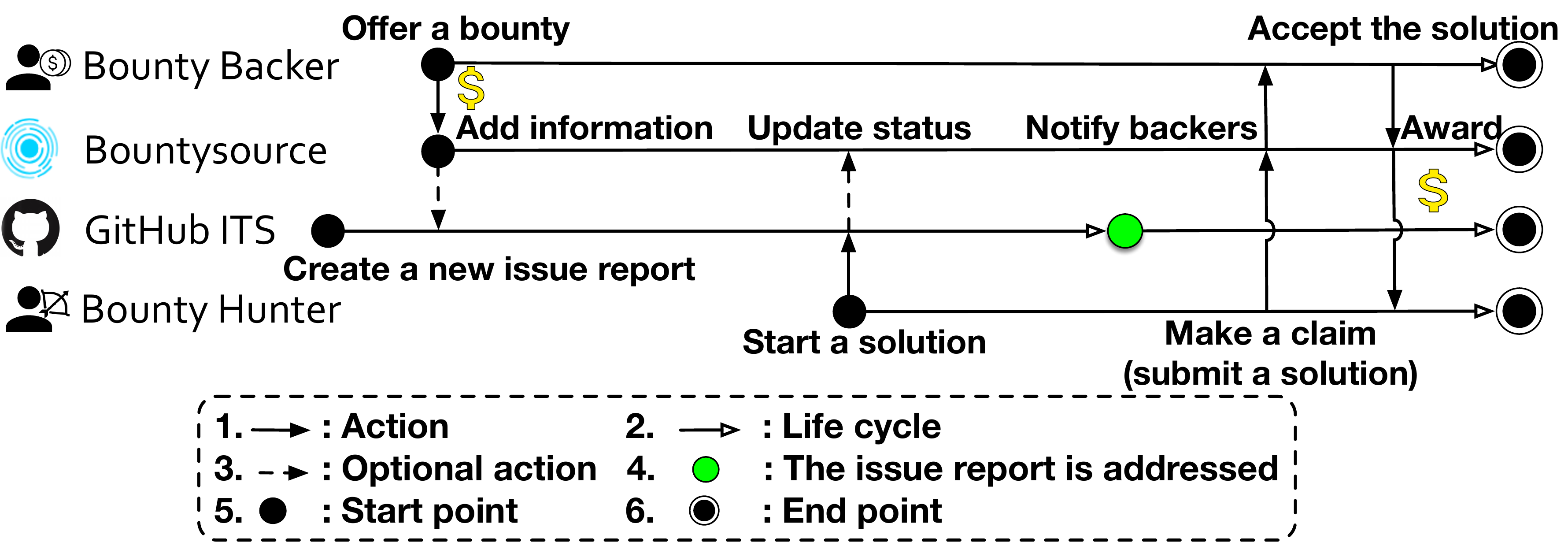}
   \vspace{-0.2in}
  \caption{The workflow of the bounty between GitHub and Bountysource.}
  \label{bountyflow}
  \vspace{-0.1in}

\end{figure}

Developers and users from more than 12 platforms (e.g., GitHub) propose bounties for issue reports through Bountysource. In this study, we focus on GitHub issue reports, since the majority of the bounties (see Section~\ref{dataset} for more details) that are proposed on Bountysource are for GitHub issue reports.
Figure~\ref{bountyflow} shows the workflow of the bounty processes between GitHub and Bountysource.
The lifecycle of a bounty starts with a bounty backer proposing a bounty for an issue report on GitHub. Bountysource will generate a link to the issue report on GitHub.
The bounty backers pledge money to Bountysource (the money is held by Bountysource) and can choose to ``advertise'' the bounty by tagging the issue report on GitHub with a bounty label (see the example\footnote{\url{http://bit.ly/2EQEA6c}} for details), appending the bounty value to the title of the issue report or mentioning the bounty in the discussion of the issue report in GitHub.
When a bounty hunter starts working on an issue, they can update their working status on Bountysource.
After the issue report is addressed, the bounty hunter can submit a claim for the bounty on Bountysource and the backer will be notified.
Once the bounty backer accepts the claim, the bounty hunter receives the money from Bountysource.

Based on the status of an issue report and whether a bounty is paid out, a bounty issue report has the following three statuses:

\noindent\textbf{Closed-paid}: the issue report is closed and the bounty has been successfully rewarded to a bounty hunter. We defined such issue reports as \emph{successful} bounty issue reports.

\noindent\textbf{Open-unpaid}: the issue report is open and the bounty is active. We defined such issue reports as \emph{failed} bounty issue reports.

\noindent\textbf{Closed-unpaid}: the issue report is closed but the bounty remains unclaimed. We defined such issue reports as \emph{ignored} bounty issue reports.

\section{Data collection}
\label{dataset}

\begin{figure}[t]
\centering\includegraphics[width=0.8\columnwidth ]{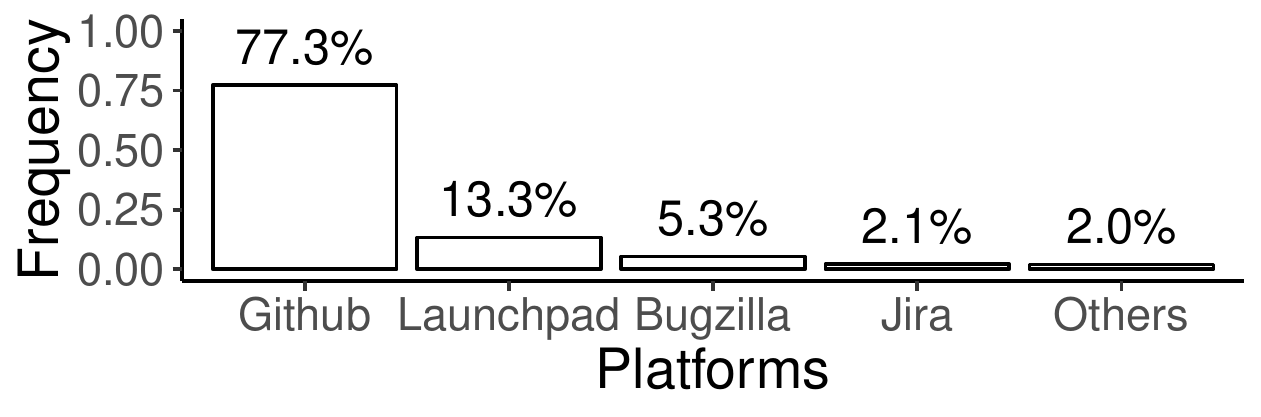}
  \caption{The distribution of Bountysource bounties across the supported ITSs.}
  \label{RatioOfIssuesITS}
  \vspace{-0.1in}

\end{figure}

In our study, we focus on the bounties that are proposed through the Bountysource platform since it is one of the most popular platforms for open source projects. As explained in Section~\ref{background}, Bountysource supports issue reports from several ITSs (e.g.,  GitHub and Bugzilla). Figure~\ref{RatioOfIssuesITS} shows the distribution of Bountysource bounties across its supported ITSs. The majority of the issue reports come from GitHub (77.3\%), hence we focus our study on the bounties that were proposed for GitHub issue reports.

\begin{figure}[t]
    \centering\includegraphics[width=9cm]{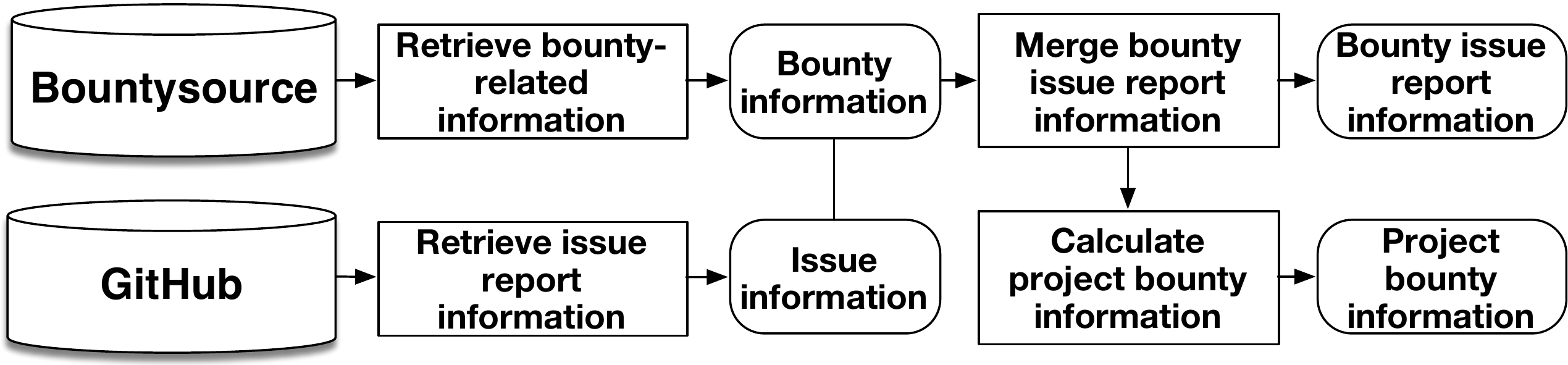}
  \caption{An overview of our data collection process.}
  \label{overview}
  \vspace{-0.1in}

\end{figure}

All information about the bounties is stored on Bountysource and all details about issue reports and their corresponding projects are stored on GitHub. Hence, we collected data for our study along three dimensions: the bounty, the issue report, and the project.

Figure~\ref{overview} presents an overview of our data collection process, which is broken down as follows:

\noindent\textbf{Step 1:} We retrieved the bounty and issue information from Bountysource using its official web API.\footnote{\url{https://bountysource.github.io/}} The bounty information includes the backers who proposed the bounty,  the proposed bounty value and the hunter who addressed the issue report. In addition, we collected basic information about the GitHub issue reports such as their id and URL.

\noindent\textbf{Step 2:} We retrieved the details of the issue reports, which are linked to Bountysource bounties by using the URL and id that we retrieved in step 1, from GitHub using its official web API.\footnote{\url{https://developer.github.com/v3/}} For example, we collected the description of the issue report, the creation date of the issue report, the comments that developers left under the issue, and the labels of the issue report.

\noindent\textbf{Step 3:} We calculated the corresponding project's bounty information for each collected bounty issue report, such as the number of total bounty issue reports of a project.

 \begin{table}[t]
   \centering
   \caption{Dataset description.}
     \begin{tabular}{p{24em}r}
    \hline
     Total number of bounties & 5,445 \bigstrut[t]\\
     Total number of claimed bounties & 2,402 \\
     Total bounty value & \$406,425 \\
     Total number of bounty hunters & 882 \\
     Total number of bounty backers & 2,534 \\
     Total number of issue reports & 3,509\\
     Total number of issue reports with multiple bounties & 795\\
     Total number of projects &  1,203 \bigstrut[b]\\
     \hline
     \end{tabular}   \label{tab:overviewOfGithub}   \vspace{-0.1in}

 \end{table}

In total, we  collected 5,445 bounties with a total value of \$406,425, together with their corresponding issue reports which were reported between Oct 19, 2012, and Oct 5, 2017. Table~\ref{tab:overviewOfGithub} gives a description of our dataset.

\section{Preliminary Study}
\label{prestudy}
\begin{figure}[t]
\centering\includegraphics[width=0.8\columnwidth ]{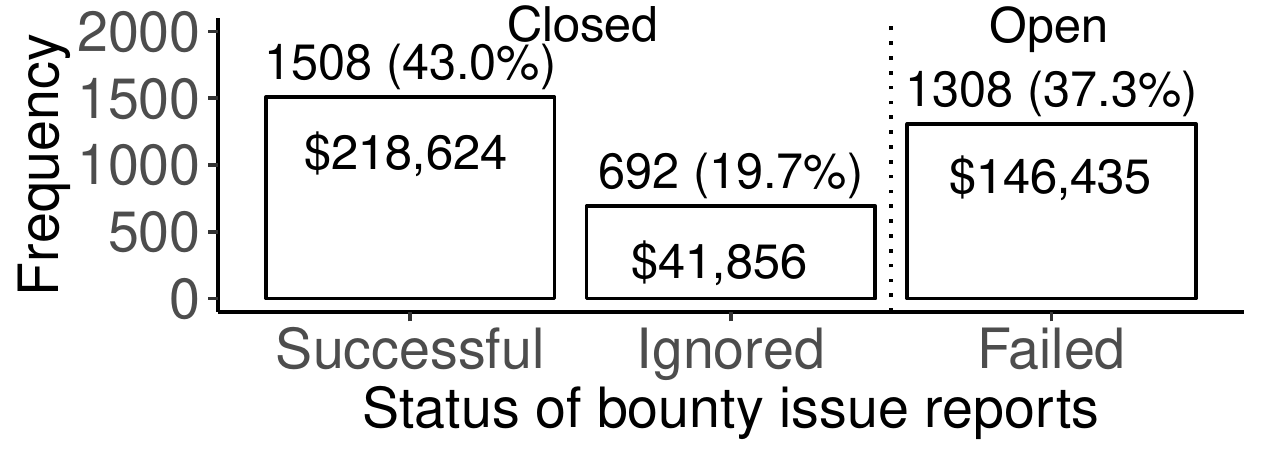}
\vspace{-0.1in}
\caption{The distribution of the possible statuses of bounty issue reports and their corresponding cumulative bounty value.}
\label{fig:pre_status_Bounties}
\vspace{-0.2in}

\end{figure}

In our preliminary study, we first present basic descriptive statistics
about the bounty: 1) the distribution of bounty issue reports across the possible statuses (i.e., successful, failed and ignored); 2) the distribution of the number of days between the reporting of the issue and its first bounty being proposed (i.e., \textit{days-before-bounty}); 3) the distribution of the frequency of bounties being used in projects (i.e., bounty-usage frequency).
From these statistics, we get a basic view of how bounties are used across projects. In addition, when a bounty issue report is closed and the bounty is paid out, we define this bounty issue report as addressed. We also calculate the issue-addressing likelihood against the bounty value and check the relationship between the issue-addressing likelihood and the bounty value.

\textbf{62.7\% of the bounty issue reports are closed, while the bounties in almost one third of these closed issue reports remain unpaid with a value of \$41,856 in total.}
Figure~\ref{fig:pre_status_Bounties} shows the distribution of bounty issue reports across the three possible statuses. 37.3\% of the bounty issue reports are failed (i.e., open-unpaid). Although 62.7\% of the bounty issue reports were closed, almost one third of their bounties were ignored (i.e., closed-unpaid). The total value of the ignored bounties (\$41,856) is ``frozen'' in the Bountysource platform unless someone claims the bounty.
In the rest of the paper, when we say the likelihood of an issue being addressed (i.e., issue-addressing likelihood), we only refer to the bounty issue reports that are successful (i.e., closed and paid out). We do not take the issue reports which were ignored into consideration because the hunters might not be driven by the bounty in such issue reports. We conducted a qualitative study of these closed-unpaid bounty issue reports to better understand them in Section \ref{rq4}.

\begin{figure}[t]
\centering\includegraphics[width=0.8\columnwidth]{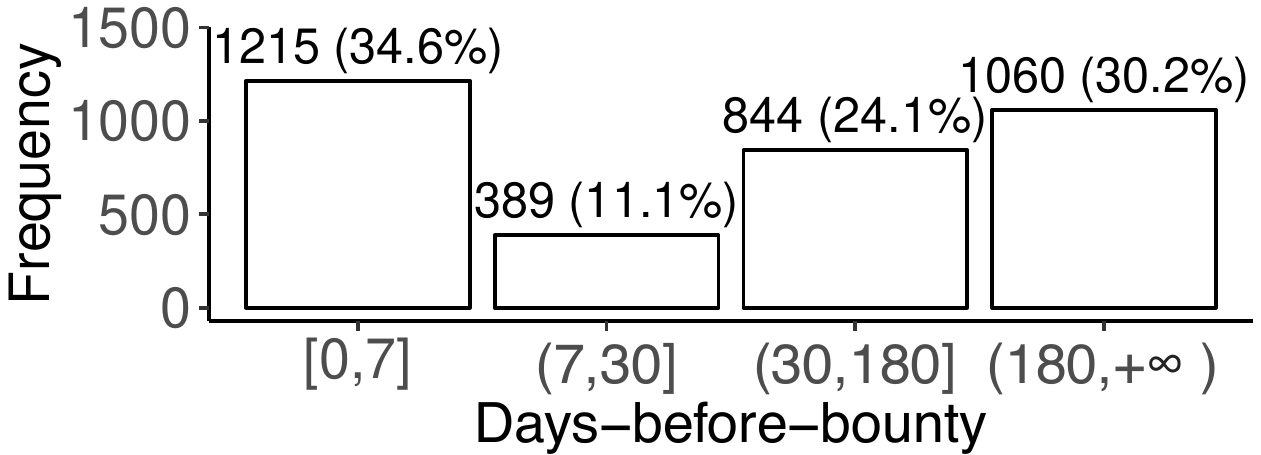}
\vspace{-0.1in}
\caption{The distribution of the number of days between the reporting of the issue and its first bounty being proposed (i.e., \textit{days-before-bounty}) in four different time ranges.}
\label{fig:pre_delta_distribution_bounty}
\vspace{-0.25in}

\end{figure}

\textbf{ 34.6\% of the bounties were proposed within 7 days from the creation of an issue report, while 30.2\% of bounties were proposed after more than 180 days.}
Figure~\ref{fig:pre_delta_distribution_bounty} shows the distribution of the \textit{days-before-bounty} metric in four continuous time ranges (i.e., 0 to 7 days, 8 days to 30 days, 31 days to 180 days, and more than 180 days). We observe that in 34.6\% of the issue reports their first bounty was proposed within 7 days since their creation, while in another 30.2\% of the issue reports their first bounty was proposed after 180 days since their creation.

 \begin{figure}[t]
   \centering
  \includegraphics[width=0.8\columnwidth]{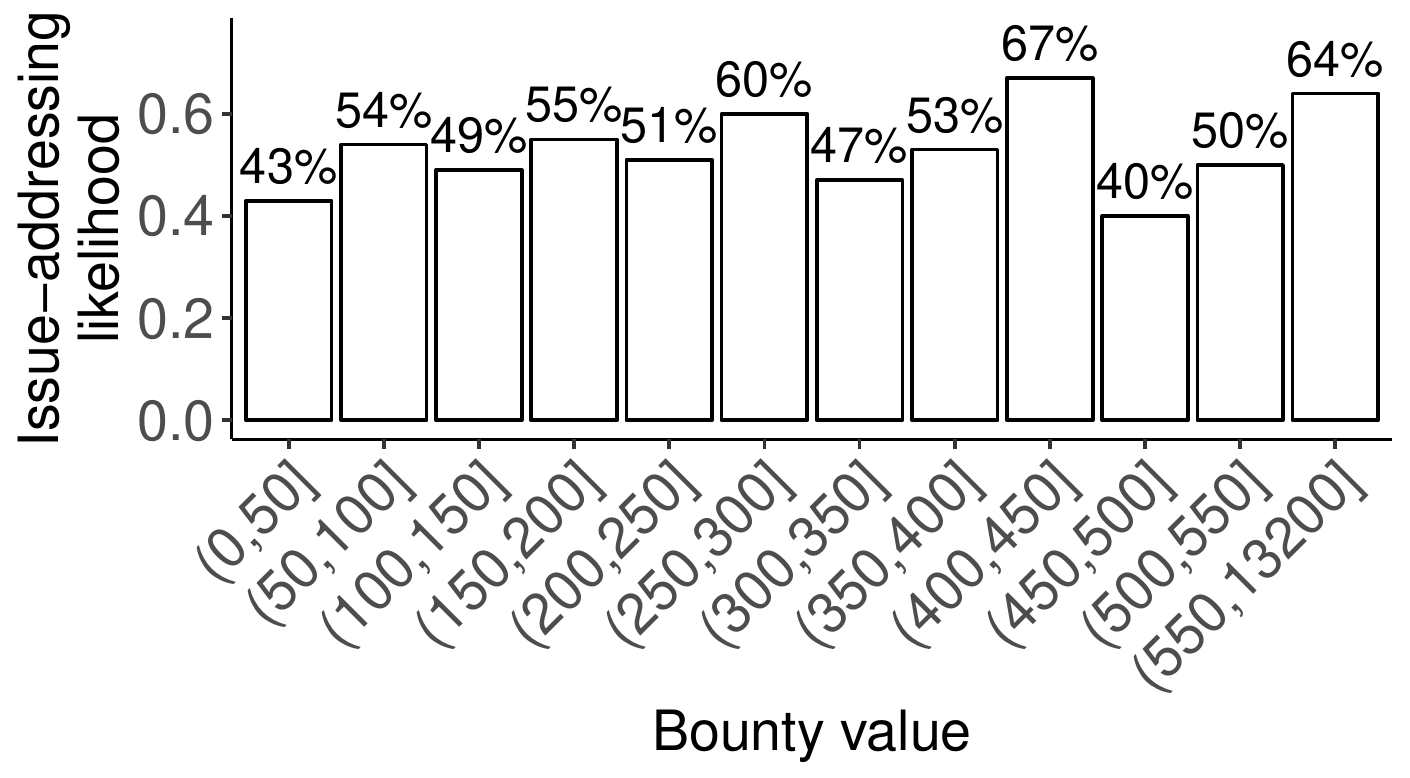}
  \caption{The issue-addressing likelihood of the proposed bounty value ranges.}
  \label{fig:pre_v_value}
  \vspace{-0.1in}

\end{figure}

\textbf{More than half of the projects only used a bounty once, while two projects used bounties very frequently (more than 100 times)}. The distribution of the bounty-usage frequency of each project is skewed (with a variance of 57.02). 768 (65\%) projects used a bounty only once, 62 projects used a bounty at least 10 times and only 9 projects used a bounty more than 50 times.

\textbf{The correlation between the bounty value and the issue-addressing likelihood is weak.} Figure~\ref{fig:pre_v_value} presents the issue-addressing likelihood of an issue report against the bounty value that an issue report has. We do not observe obvious patterns between them. The correlation between the bounty value and the issue-addressing likelihood is 0.14, which indicates a weak correlation.

\rqbox{
Counter-intuitively, the correlation between the bounty value and the issue-addressing likelihood is weak. However, this low correlation may be due to the variation of the frequency of bounties that were used before in different projects and the timing of proposing the bounty.
}
Therefore, in Section~\ref{rq1} we investigate how the issue-addressing likelihood changes across projects with different bounty-usage frequencies. We study the relationship between the timing of proposing bounties and the issue-addressing likelihood in Section~\ref{rq2}.

\section{Studying how the issue-addressing likelihood changes across projects which have different bounty-usage frequencies}

\label{rq1}

Bounties are frequently used in some projects while rarely used in others (see Section~\ref{prestudy}). The frequency of using bounties in a project may reflect the degree of experience that the project has with bounties and may have an impact on the issue-addressing process.
Therefore, in this RQ, we investigate how the issue-addressing likelihood changes across different projects with different bounty-usage frequencies. Furthermore, we study the relationship of the issue-addressing likelihood with the bounty value and the project activities (e.g., creating pull requests).
With a better understanding of the bounty across different projects, we can provide insights on how to better use bounties.

\begin{figure}[t]
\centering\includegraphics[width=\columnwidth ]{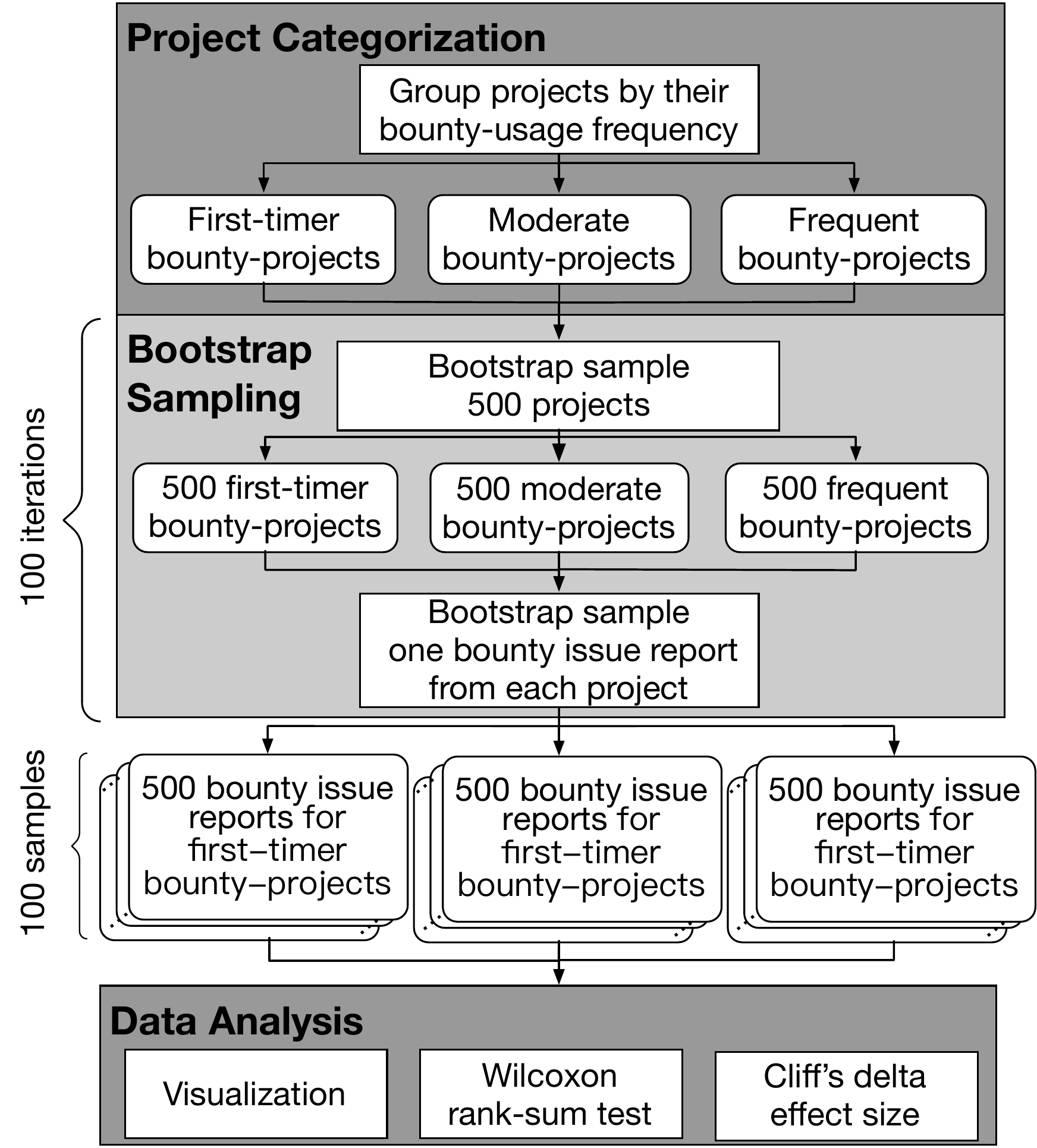}
\caption{An overview of our approach. }
\label{fig:rq1_approach}
\vspace{-0.2in}

\end{figure}

\noindent\textbf{Approach:} Figure~\ref{fig:rq1_approach} gives an overview of our approach. We elaborate on each step below.

\noindent\emph{\textbf{Project categorization:}} Given the variance (see Section~\ref{prestudy}) of the bounty-usage frequency for different projects, it is not advisable to study all the issue reports as one group when we study the bounties at the issue-level. Therefore we categorize the projects into the following three groups:

\begin{enumerate}
    \item \textbf{First-timer bounty-projects:} Projects which have only one bounty issue report.
    \item \textbf{Moderate bounty-projects:} Projects which have 2 to 50 bounty issue reports.
    \item \textbf{Frequent bounty-projects:} Projects which have more than 50 bounty issue reports.
\vspace{-0.01in}
\end{enumerate}

It is important to study the bounties in the first-timer bounty-projects, since users of such projects may not have former bounty experience. Insights on the usage (e.g., how large of a bounty should be proposed?) and the impact (e.g., the issue-addressing likelihood) of bounties in these projects may benefit other projects which are new to using bounties. We grouped the projects that have more than 50 bounty issue reports as well since we assume that in such projects the community is more familiar with the use of bounties.
After grouping the projects into the above mentioned three groups we ended up with 768 (65\%) first-timer bounty-projects, 400 (34\%) moderate bounty-projects, and 9 (1\%) frequent bounty-projects.

\noindent\emph{\textbf{Bootstrap sampling:}} After grouping the projects into the three groups, we used a bootstrap sampling approach to sample issue reports across projects. The purpose of using bootstrap sampling is to reduce the bias that is caused by the unbalanced number of projects across the three groups. We first randomly sampled 500 projects from each group with replacement. Then we randomly sampled one bounty issue report from each sampled project. The purpose of sampling one report from each project is to avoid a bias towards the projects that have more issue reports than other projects in the same group. Aa a result, we sampled 500 bounty issue reports from each of the 3 project groups. To make our results stable and reliable, we repeated the above sampling process 100 times with different random seeds. We ended up with 100 samples with 1,500 issue reports each (i.e., 500 issue reports for each group).

\noindent\emph{\textbf{Data Analysis:}} For each sample, we calculated the issue-addressing likelihood across the three project groups and visualized the results in plots. To compare the differences between two data distributions (e.g., the differences of the bounty values between successful and failed bounty issue reports), we used the Wilcoxon rank-sum test~\cite{bauer1972constructing}, which does not require the sample to be normally distributed. We consider the two distributions as significantly different when the p-value of the test is smaller than 0.05. Furthermore, we applied Cliff's delta $d$ \cite{long2003ordinal} effect size to quantify the magnitude of the difference. We use the following thresholds for $d$~\cite{romano2006appropriate}: $ \vert d \vert \leq$0.147 (negligible); 0.147 $ < \vert d \vert \leq $0.33 (small); 0.33 $ < \vert d \vert \leq $0.474 (medium); 0.474 $ < \vert d \vert  \leq $1 (large).

\begin{figure}[t]
    \centering
    \begin{subfigure}[t]{0.45\columnwidth }
        \centering
        \includegraphics[width=\linewidth]{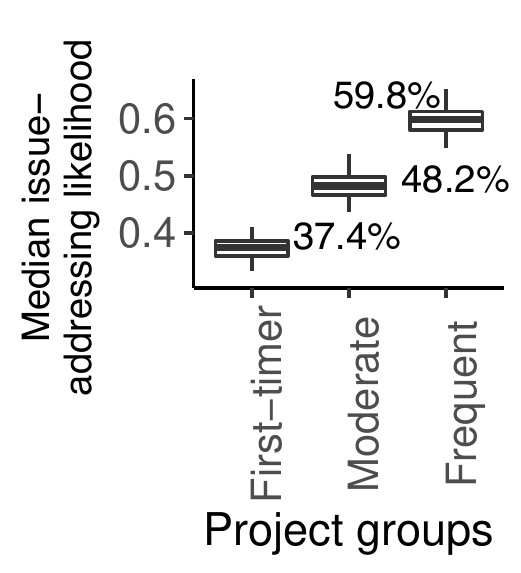}
        \caption{}
		\label{fig:rq1_likelihood_ratio500}
    \end{subfigure}    ~
    \begin{subfigure}[t]{0.45\columnwidth}
        \centering
        \includegraphics[width=\linewidth]{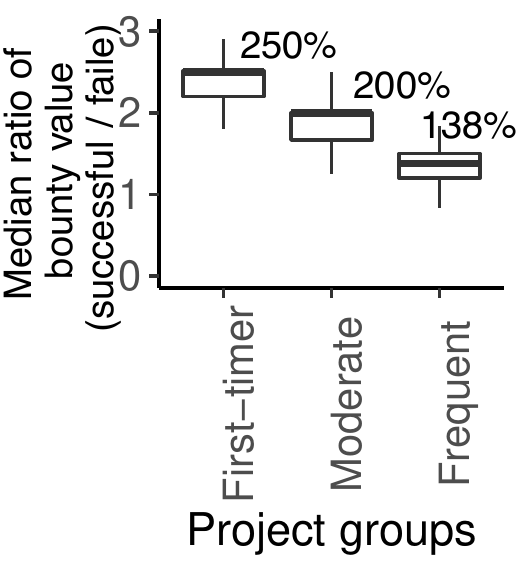}
        \caption{}
          \label{fig:rq1_ratio_cp_ou500}
    \end{subfigure}
    \vspace{-0.1in}
    \caption{The distribution of the median issue-addressing likelihood (a) and the distribution of the median ratio of the bounty value of successful bounty issue reports to the value of failed bounty issue reports (b) for each project group for 100 samples.
    }
    \vspace{-0.2in}
\end{figure}

\noindent\textbf{Results:}
\textbf{Issue reports in projects which have a higher bounty-usage frequency are more likely to be addressed.} Figure~\ref{fig:rq1_likelihood_ratio500} shows the median issue-addressing likelihood for each project group. We observed a positive relationship between this likelihood and the bounty-usage frequency, which indicates an issue report is more likely to be addressed in a project with a higher bounty-usage frequency. Our statistical test results show that the issue-addressing likelihood is significantly different across these three groups of projects with a large effect size. One possible explanation is that the lower bounty-usage frequency of a project may indicate that the backers have less experience in proposing a bounty (e.g., at the proper time with a proper value) and the hunters react to bounties less actively than in projects with a higher bounty-usage frequency.

\textbf{The successful bounty issue reports have a relatively higher bounty value than failed bounty issue reports, particularly in the first-timer bounty-projects.}
Figure~\ref{fig:rq1_ratio_cp_ou500} shows that the median ratio (the bounty values of successful reports compared to those of failed reports) is larger than 1 in all 3 project groups. In other words, the successful bounty issue reports have higher bounty values than the failed bounty issue reports. In addition, comparing the ratio across project groups, the first-timer bounty-projects have the largest ratio (2.5) among the three groups, indicating that developers may want more money to address an issue in first-timer bounty-projects than in other projects. Our statistical test results shows that the differences in bounty value between successful and failed bounty issue reports are significant (p-value $<$ 0.05) in the first-timer bounty-projects for all samples with a non-negligible Cliff's delta effect size (72\% samples have a small effect size and 28\% samples have a medium effect size).
In 97\% of the samples of the moderate bounty-projects the bounty value of successful issue reports is significantly higher than that of failed issues (with a non-negligible effect size in 77\% of the cases).
For the frequent bounty-projects, in 92\% of the samples the differences were not statistically significant.

\begin{figure}[t]
\centering\includegraphics[width=\columnwidth]{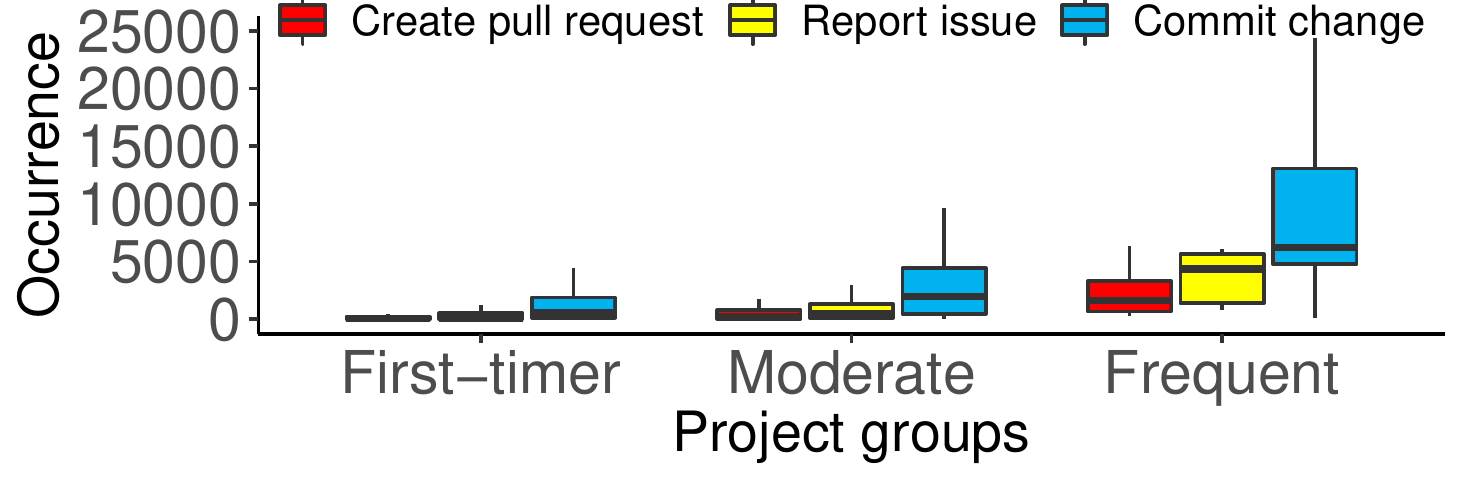}
\caption{The distributions of the occurrences of three activities (i.e., the create pull request, the report issue and the commit change) in each project group.}
\label{fig:rq1_activity}
\vspace{-0.2in}

\end{figure}

One possible explanation is that first-timer bounty-projects may not be as active as moderate and frequent projects. Therefore, backers would be required to propose bounties with higher values to attract enough attention from the community for addressing issues. To investigate this explanation, we examined the activity of the projects in terms of the number of pull requests, issue reports, and commits. Figure~\ref{fig:rq1_activity} shows the distributions of the occurrences of these three activities in each project group.
\textbf{Projects with fewer bounty issue reports are usually less active (in terms of the number of pull requests, issue reports, and commits) than projects with more bounty issue reports.} Another possible explanation is that the backers in the first-timer bounty-projects have no experience in proposing bounties and sometimes overestimate the value of addressing an issue report. In this situation, the overestimated bounty issue reports may attract more attention from the community due to their ``easy money'' and get addressed easier.

\rqbox{The issue-addressing likelihood is higher in projects which used bounties more frequently. Successful bounty issue reports have higher bounty values than failed bounty issue reports in the projects which used bounties less frequently.}

\section{Studying the relationship between the timing of proposing a bounty and the issue-addressing likelihood}
\label{rq2}

In Section~\ref{prestudy}, we observe different patterns for the timing of proposing bounties, which may have different relationships with the issue-addressing likelihood. In this section, we investigate the relationship between the timing of proposing bounties and the issue-addressing likelihood. With a better understanding of the relationship, we can provide insights into how to improve the timing of proposing a bounty.

\textbf{Approach:}
We use the same data (100 samples of 1,500 issue reports for each sample) as RQ1. For each sample, we grouped data into four timing ranges based on the number of elapsed days between the creation of an issue report and the proposal of its first bounty(i.e., \textit{days-before-bounty}). We defined the four timing ranges as follows:
\begin{enumerate}
    \item \textbf{[0,7]:} the first bounty was proposed within seven days after the issue was reported.
    \item \textbf{(7,30]:} the first bounty was proposed between 7 and 30 days after the issue was reported.
    \item \textbf{(30,180]:} the first bounty was proposed between 30 and 180 days after the issue was reported.
    \item \textbf{(180,$\infty$):} the first bounty was proposed after 180 days after the issue was reported.
\end{enumerate}

Then we calculate the issue-addressing likelihood across the timing ranges for each sample to study the relationship between the timing and issue-addressing likelihood.
To further study the relationship between the issue-addressing speed and the timing, we also calculate the time it takes to close the issue report after the first bounty is proposed (i.e., \textit{time-to-close}) in each timing range for each sample.
Furthermore, we also study the timing in terms of different project groups for each sample.

\begin{figure}[t]
    \centering
    \begin{subfigure}[t]{0.5\columnwidth }
        \centering
        \includegraphics[width=\linewidth]{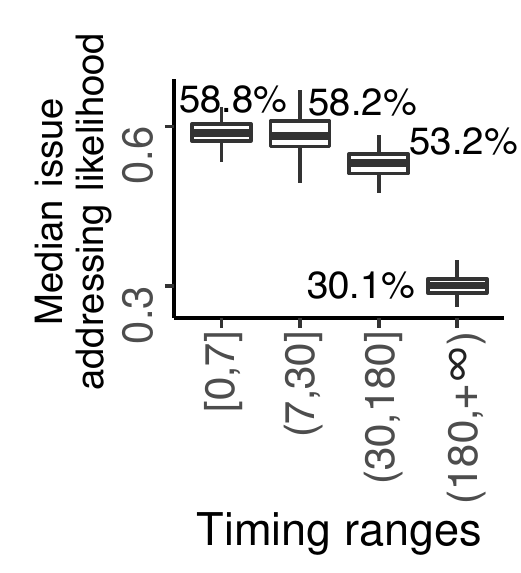}
        \caption{}
		\label{fig:rq2_delta_ratio}
    \end{subfigure}    ~
    \begin{subfigure}[t]{0.5\columnwidth}
        \centering
        \includegraphics[width=\linewidth]{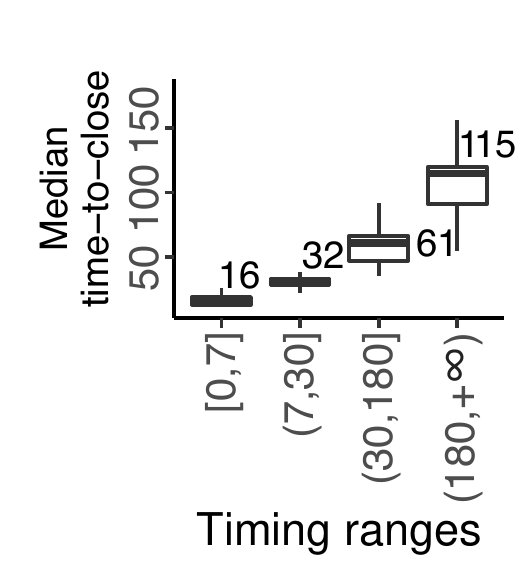}
        \caption{}
          \label{fig:rq2_closeday_deltaday}
    \end{subfigure}
    \vspace{-0.1in}
    \caption{The distribution of the median issue-addressing likelihood (a) and the  distribution of the median time to close an issue report (i.e., \textit{time-to-close}) (b) across the  timing ranges for the 100 studied samples.
    }
\end{figure}
\begin{figure}[t]
\centering\includegraphics[width=0.8\columnwidth]{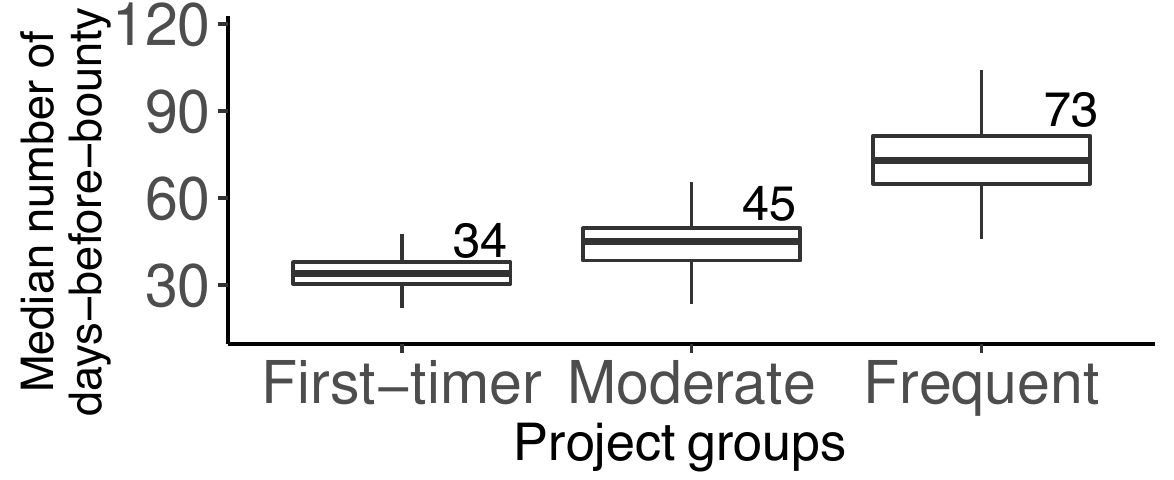}
  \caption{The distribution of the median number of days between the creation of the issue report and the proposed of the first bounty to the bounty is proposed (i.e., \textit{days-before-bounty}) across different project groups for 100 samples.}
  \label{fig:rq2_medianDeltaTime}
  \vspace{-0.15in}

\end{figure}

\noindent\textbf{Results:}\textbf{ In general, issue reports for which bounties were proposed earlier have a higher likelihood of being addressed.} Figure~\ref{fig:rq2_delta_ratio} shows the distribution of the median issue-addressing likelihood across the timing ranges over 100 samples. We calculated the Wilcoxon rank-sum test and Cliff's delta test to measure the differences between two distributions. The results show that there is no significant difference between the distributions of ``[0,7]'' and ``(7,30]'' while there exist significant differences with large effect sizes for the other two distributions.
The likelihood is getting smaller as the \textit{days-before-bounty} gets larger, especially for the issue reports in which bounties are proposed after 180 days where the likelihood drops to 30\%.

One possible explanation is that as time progresses, the risk of a report becoming obsolete exists, leaving the issue report unaddressed even after a bounty is proposed.
For example, an issue report\footnote{\url{https://github.com/bhdouglass/uappexplorer/issues/69}} that was created on Feb 4, 2016 in the \code{uappexplorer} project requested a new feature for an Ubuntu Phone Application.
The owner of the application and another developer both showed great interest in this issue. Because of the lack of time, the feature was never added. A bounty of \$5 was proposed\footnote{\url{http://bit.ly/2Q3BIns}} after almost one year (i.e., on Jan 12, 2017).
However, the issue report was closed because Ubuntu Phone was no longer used making the issue report obsolete. Because of the low value, the bounty was not refundable. In other words,\textbf{ backers carry the risk of wasting their money by proposing small bounties on such long-standing issue reports.}

Another possible reason for the lower issue-addressing likelihood of the issue reports for which bounties were proposed later is the potential difficulty of such issue reports.
Figure~\ref{fig:rq2_closeday_deltaday} shows the median time that was taken to close issues (i.e., \textit{time-to-close}) across each timing range. We observe that the issue reports in which bounties were proposed later took more time to be addressed.

\textbf{Backers proposed bounties earlier in the first-timer bounty-projects.}
Figure~\ref{fig:rq2_medianDeltaTime} presents the distribution of the median value of the \textit{days-before-bounty} metric across the project groups for each sample. The median \textit{days-before-bounty} is 34, 45 and 73 for each project group, respectively.
We calculated the Wilcoxon rank-sum test and Cliff's delta $d$ effect size to measure the differences of distributions between the first-timer bounty-projects and the moderate bounty-projects, the moderate bounty-projects and the frequent bounty-projects, respectively. The result shows that any two pairs of distributions are significantly different with large Cliff's delta effect sizes, indicating that the \textit{days-before-bounty} is higher in projects with a higher bounty-usage frequency.
One explanation is that the activity of first-timer bounty-project is lower (see Figure~\ref{fig:rq1_activity}), which may encourage backers to propose a bounty earlier to attract developers to get an issue addressed.

\rqbox{In general, issue reports for which bounties were proposed earlier have a higher likelihood of being addressed. In addition, the risk of losing money exists for backers who propose small bounties for long-standing issue reports.
}

\section{Using a logistic regression model to study the relationship between the studied factors and the issue-addressing likelihood}

\label{rq3}

In the previous sections, we studied the relationship between the issue-addressing likelihood, the bounty-usage frequency and the timing of proposing a bounty. However, there may exist more factors that potentially affect the issue-addressing likelihood. In this section, we conduct a deeper study on the relationship between other factors and the issue-addressing likelihood.
With a better understanding of this relationship, we can provide insights into how to use a bounty to improve the issue-addressing process.

\begin{figure*}[t]
\centering\includegraphics[width=17cm]{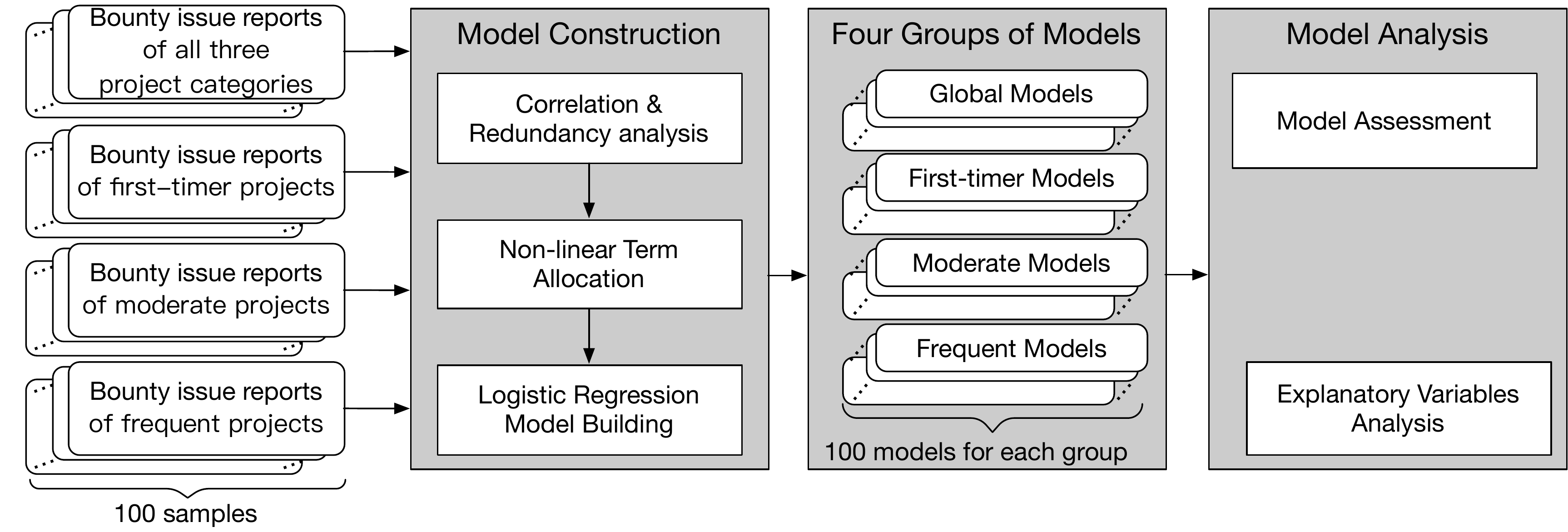}
\vspace{-0.2in}
\caption{An overview of the model construction and analysis.}
\label{fig:rq3_flow}
\vspace{-0.1in}
\end{figure*}

\subsection{Approach}
In this section, we use the same data as in Section~\ref{rq1} to build logistic regression models to investigate the relationship between the studied factors and the issue-addressing likelihood.
Firstly, we build models on the entire set of bounty issue reports to understand the global relationship (referred to as the \textbf{global model}). Secondly, as we can see in Section~\ref{rq1}, the issue-addressing likelihood changes across project groups (i.e., the first-timer bounty-projects, the moderate bounty-projects, and the frequent bounty-projects). To understand the relationship within each project group, we build logistic regression models on the bounty issue reports of each group separately. To condense our writing, we refer to the model for the first-timer, moderate, and frequent bounty-projects as the \textbf{first-timer model}, \textbf{moderate model}, and \textbf{frequent model}, respectively.
We use the logistic regression modeling technique since it is a robust and highly interpretable technique, which has been applied successfully in software engineering before~\cite{wang2017understanding,mcintosh2016empirical}.
Figure~\ref{fig:rq3_flow} gives an overview of our approach.
Below, we elaborate on the studied factors, the processes of the model constructions, and the analysis of our models.

\subsubsection{Studied factors}
\begin{table*}[]
\centering
\footnotesize
\caption{The description of and rationale for the factors in the \emph{Issue report basic} and the \emph{Issue report bounty} dimensions. The factors which are marked with `*' are  time-dependent factors which are calculated at the time when the bounty is proposed}
\label{tab:factors1}

     \begin{tabular}{p{12em}p{29em}p{19em}}
          \toprule
     \multicolumn{1}{l}{\textbf{Factor name}}  & \textbf{Description} & \multicolumn{1}{p{19em}}{\textbf{Rationale}} \\

     \midrule
     \multicolumn{3}{p{30em}}{\textbf{Issue report basic}}\\
     \midrule
      I\_content\_len*                  & The length of an issue report and its comments (in characters).  &\multirow{5}[-3]{19em}{\parbox{19em}{These factors reflect the amount of supportive information that an issue report has. Issue reports with more supportive information may help developers to address them.}} \\
     I\_code\_len* & The total length of the code snippets in an issue report and its comments (in characters). &  \\
      I\_code\_proportion*              & The proportion of code in an issue report and comments (i.e., $\frac{I\_code\_len}{I\_content\_len}$).\makecell{ \rule{0pt}{12pt} } &  \\
      \midrule
    I\_link\_cnt*                     & The number of links in an issue report and its comments. & \multirow{5}[-4]{19em}{\parbox{19em}{The discussion activities reflect the popularity of an issue report, which may have a relationship with the issue-addressing likelihood.}} \\
     I\_img\_cnt*                      & The number of images in an issue report and its comments. &  \\

     I\_cmnt\_cnt                     & The number of comments that an issue report received. & \\
	I\_participant\_cnt* & The number of participants in the discussion of an issue. & \\
     I\_cmnt\_per\_day\_mean* & The mean number of comments per day for an issue report. &  \\
 \midrule
     \textbf{Issue report bounty} & \multicolumn{2}{l}{\textbf{Description (d) - Rationale (r)}} \\
     \midrule
     I\_B\_days\_before\_bounty*  & \multicolumn{2}{p{48em}}{d: The number of days between the creation of an issue report and its first bounty.\newline r: Different timing of proposing bounties have relationship with the issue-addressing likelihood.}      \\
                I\_B\_total\_value & \multicolumn{2}{p{48em}}{d: The total bounty value of the issue report. \newline r: A higher bounty may attract more developers.} \\
                I\_B\_cnt    &  \multicolumn{2}{p{48em}}{d: The number of bounties that a bounty issue report has. \newline r: A higher number indicates that more backers are interested in getting this issue  addressed.} \\
                I\_B\_has\_label  &  \multicolumn{2}{p{48em}}{d: Whether a bounty issue report is tagged with a bounty label. \newline r: A bounty label could help draw attention from the community, which may impact the issue-addressing likelihood. } \\
                I\_B\_timing\_range*   & \multicolumn{2}{p{48em}}{d: The range of the timing of proposing the first bounty.                                                                                 \newline r: The timing of proposing a bounty has a relationship with the issue-addressing likelihood (see Section~\ref{rq2}).} \\
     \bottomrule
          \end{tabular} \end{table*}
\begin{table*}[ ]
\centering
\caption{The description of and rationale for the factors in the \emph{Project bounty} and the \emph{Backer experience} dimensions. The factors which are marked with `*' are  time-dependent factors which are calculated at the time when the bounty is proposed}
\label{tab:factors2}

     \begin{tabular}{p{10em}p{29em}p{21em}}
     \toprule
     \multicolumn{1}{l}{\textbf{Factor name}}  & \textbf{Description} & \multicolumn{1}{p{21em}}{\textbf{Rationale}} \\
     \midrule
     \multicolumn{3}{p{30em}}{\textbf{Project bounty}}\\
     \midrule
     P\_B\_I\_cnt*                     &  The total number of issue reports with at least one bounty of a project.  & \multicolumn{1}{r}{\multirow{5}[-5]{21em}{\parbox{21em}{These five factors reflect the bounty activity of the project. A different level of activity may have a different impact on the issue-addressing likelihood in the project.}}} \\
     P\_B\_paid\_cnt* & The total number of paid bounty issue reports of a project. &  \\
     P\_B\_open\_cnt*     &  The number of open bounty issue reports of a project.  &\\
     P\_B\_paid\_proportion* & The proportion of paid bounty issue reports of a project. &  \\
     P\_B\_total\_value* & The total value of the bounties of a project. &  \\
\midrule
      P\_B\_usage\_group                    &   The group of projects.                                                                                  & \multicolumn{1}{p{21em}}{Different groups of projects have different issue-addressing likelihoods (see Section~\ref{rq1}).} \\
     \midrule
     \multicolumn{3}{p{30em}}{\textbf{Backer experience}}\\
     \midrule
     Backer\_exp\_B\_median/sum/max\_value* & The median/sum/max value of bounties which the backers of this bounty have ever proposed in the past. & \multicolumn{1}{r}{\multirow{2}[1]{21em}{\parbox{21em}{Bounties from a backer who has proposed bounties often, or proposed high-value bounties in the past may attract more attention from developers.}}} t\\
     Backer\_exp\_B\_median/sum/max\_cnt* & The median/sum/max number of bounties which the backers of this bounty have ever proposed in the past. &  \\
     \midrule
     Backer\_role\_any\_insider* & Whether any of the backers who has ever contributed to the project.  & \multicolumn{1}{r}{\multirow{2}[1]{21em}{\parbox{21em}{A backer who has ever interacted with the project before may help the bounty attract more attention from the community.}}} \bigstrut\\
Backer\_role\_have\_reporter* & Whether the issue reporter is one of the backers for that issue report.\\
     \midrule

     \end{tabular}\vspace{-0.2in}
 \end{table*}
We consider 27 factors along 4 dimensions:
\begin{enumerate}
    \item \textbf{Issue report basic:} Eight factors which can estimate the length and the popularity of an issue report.
    \item \textbf{Issue report bounty:} Five factors which describe the bounty usage within a bounty issue report.
    \item \textbf{Project bounty:} Six factors which reflect the bounty usage within a project.
    \item  \textbf{Backer experience:}  Eight factors which capture the bounty experience of the backers of a bounty issue report.
\vspace{-0.01in}
\end{enumerate}

Table~\ref{tab:factors1} and Table~\ref{tab:factors2} summarize the description of and rationale behind the studied factors. The factors which are marked with `*' are time-dependent factors which are calculated at the time when the bounty is proposed.
Note that the factors in the project bounty, issue report basic, and backer experience dimensions cannot be changed by a backer who wants to propose a bounty. Hence we include these factors as control factors in our models.

\subsubsection{Model construction}
Similar to prior studies~\cite{gopi:2017,wang2017understanding,mcintosh2016empirical,kabinna2018examining}, we first removed correlated and redundant factors by using the Spearman rank correlation test and the redundancy analysis to avoid multicollinearity. We ended up with three factors in the project bounty dimension, six factors in the issue report basic dimension, four factors in the issue report bounty dimension, and three factors in the backer experience dimension.
Then we added non-linear terms (i.e. NL) in the model to capture the more complex relationship in the data by employing restricted cubic splines~\cite{Harrell:2006}.
Finally, we used the \code{rms} \code{R} package \footnote{\url{https://cran.r-project.org/web/packages/rms/index.html}} to implement our logistic regression models (i.e., the first-timer, moderate, frequent, and global models) based on 100 samples and ended up with 400 models. See our appendix~\cite{appendix} for more details about our model construction.

\subsubsection{Model analysis}
\begin{figure}[t]
\centering\includegraphics[width=0.7\columnwidth ]{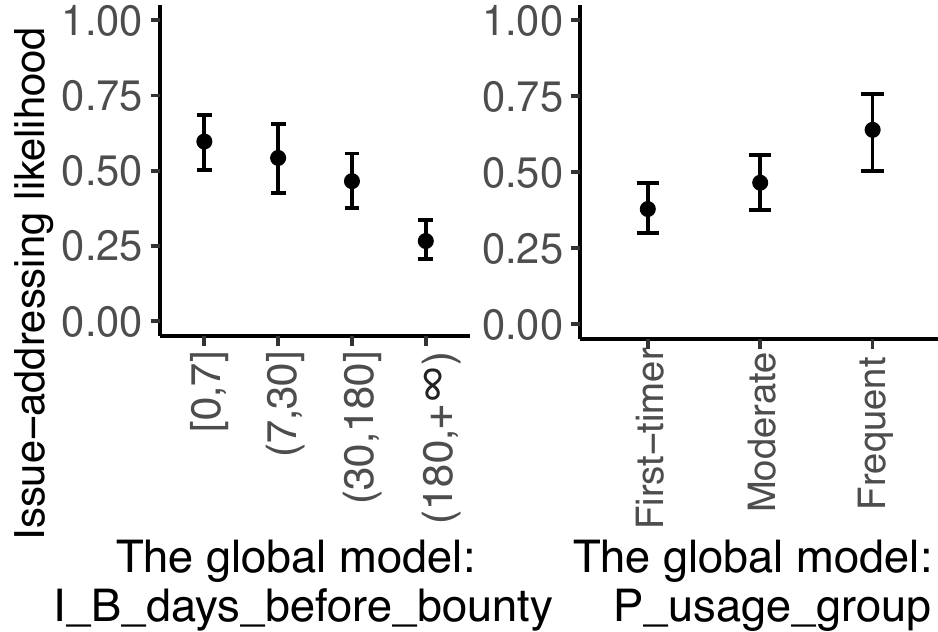}
\caption{The plots show the relationship between the studied factors and the issue-addressing likelihood in the global models in the selected sample. For each plot, we adjusted all factors except the studied factor to their median value in the model and recomputed the issue-addressing likelihood. The grey area represents the 95\% confidence interval.}
\vspace{-0.1in}
\label{fig:rq3_trend_global}
\end{figure}

\begin{figure}[t]
\centering\includegraphics[width=0.8\columnwidth ]{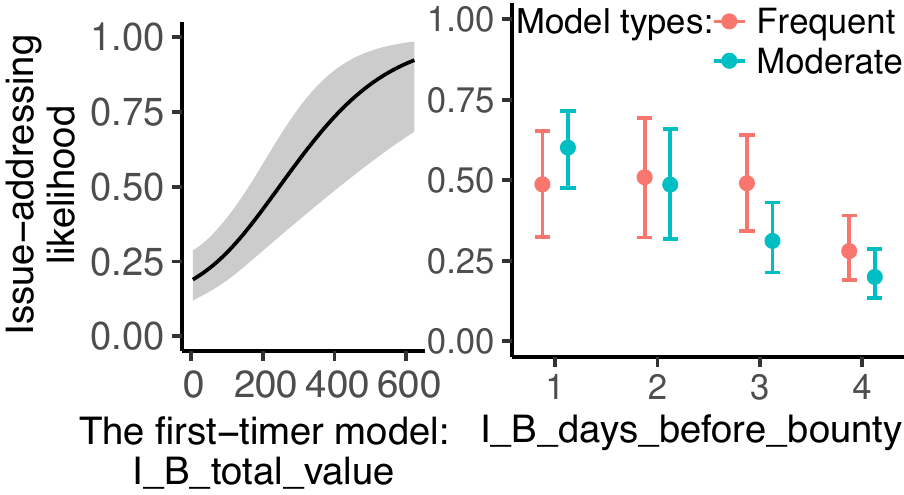}
\caption{The plots show the relationship between the studied factors and the issue-addressing likelihood in the first-timer, moderate and frequent models in the selected sample. For each plot, we adjusted all factors except the studied factor to their median value in the model and recomputed the issue-addressing likelihood. The grey area represents the 95\% confidence interval.}
\vspace{-0.1in}
\label{fig:rq3_trend_two}
\end{figure}

For each group of logistic regression model, we used the Area Under the ROC Curve (i.e., \textit{AUC}) and a bootstrap-derived approach \cite{efron1986biased} to assess the explanatory power of models following prior studies~\cite{mcintosh2016empirical,wang2017understanding,kabinna2018examining}. The AUC ranges from 0 to 1 (0.5 is the performance of a random guessing model) and a higher AUC means that the model has a higher ability to capture the relationships between explanatory factors and the response factor. The \textit{optimism value} is calculated by the bootstrap-derived approach and the small optimism value indicates that the model does not suffer from overfitting.
To study the impact of each factor on the issue-addressing likelihood. We used the \textbf{anova} function in the R \textbf{rms} package to compute the Wald $\chi^2$ value which reveals the impact.
The larger the Wald $\chi^2$ value of a factor is, the larger impact the factor has on the issue-addressing likelihood.
We computed the Wald $\chi^2$ value for each factor for each model and used the median Wald $\chi^2$ value of each factor within a group to represent the impact of that factor in that group.
In addition, to further understand how a factor influences the value of the response variables, we used the \textbf{Predict} function in the \textbf{rms} R package to plot the estimated issue-addressing likelihood against a factor. Since all models across 100 samples showed similar patterns of influence for the factors, we randomly selected a sample as an example to build models and visualize the results (see Figure~\ref{fig:rq3_trend_global} and Figure~\ref{fig:rq3_trend_two}).
The analysis allows us to further understand how a factor affects the issue-addressing likelihood. For the detailed results of our model analysis, such as the $\chi^2$ values, we refer the reader to our online appendix~\cite{appendix}.

  \begin{table}[t]
   \centering
   \caption{The top two most important factors, the median AUC (i.e., M\_A) and the median optimism value (i.e., M\_O) for our four group of models.}
     \begin{tabular}{lp{8em}ll}
     \toprule
     Model group & \multicolumn{2}{c}{\rule{-14pt}{0pt} Factor ranking} & \rule{-11pt}{0pt}M\_A/M\_O \bigstrut\\
     \midrule
           & \multicolumn{1}{c}{\rule{-13pt}{0pt}1} & \multicolumn{1}{c}{\rule{-14pt}{0pt} 2} &  \bigstrut\\
     \midrule
     Global  &\rule{-13pt}{0pt} I\_B\_timing\_range &\rule{-14pt}{0pt} P\_B\_usage\_group &\rule{-11pt}{0pt} 0.72/0.02 \bigstrut\\
     \midrule
First-timer &\rule{-13pt}{0pt} I\_B\_total\_value &\rule{-14pt}{0pt} I\_B\_timing\_range &\rule{-11pt}{0pt} 0.74/0.03 \bigstrut\\
     \midrule
  Moderate &\rule{-13pt}{0pt} I\_B\_timing\_range &\rule{-14pt}{0pt} I\_code\_proportion &\rule{-11pt}{0pt} 0.71/0.04 \bigstrut\\
     \midrule
   Frequent &\rule{-13pt}{0pt} I\_B\_timing\_range &\rule{-14pt}{0pt} P\_B\_paid\_proportion &\rule{-11pt}{0pt} 0.81/0.03 \bigstrut\\
     \bottomrule
     \end{tabular}   \label{tab:top3impFactors} \end{table}

\subsection{Results}
\textbf{Our models explain our dataset well and have a reliable performance.} The median AUCs for each group of models are all at least 0.71 (see Table~\ref{tab:top3impFactors}), which indicates that our models explain the dataset well and the low median optimism values (between 0.02 and 0.04) indicate that our models do not overfit the dataset.

\textbf{In the global view, the timing of proposing the bounties and the bounty-usage frequency are the top two most important factors that impact the issue-addressing likelihood.}  Table~\ref{tab:top3impFactors} shows the top two important  factors (ranked by median Wald's $\chi^2$ value) in the global models across 100 samples.
\textit{I\_B\_timing\_range} (i.e., the range of the timing of proposing bounties) and \textit{P\_B\_usage\_group} (i.e., the group of projects' bounty-usage frequency) are the most important factors which contribute the most explanatory power to the models. This observation echoes our findings in Section~\ref{rq1} and ~\ref{rq2} that the timing of proposing bounties and the bounty-usage frequency have a strong impact on the issue-addressing likelihood.

Figure~\ref{fig:rq3_trend_global} shows the relationship between the issue-addressing likelihood and the top two most important factors for the global model. \textit{I\_B\_timing\_range} has a negative relation with the issue-addressing likelihood, which indicates that issue reports for which bounties are proposed earlier have a higher likelihood of being addressed. This observation echoes with our findings in Section~\ref{rq2}.

\textbf{In the projects that use bounties moderately and frequently, issue reports are more likely to be addressed if backers propose bounties on an issue report earlier.} Table~\ref{tab:top3impFactors} shows that \textit{I\_B\_timing\_range} is the most important factor in the moderate and the frequent models. Especially in the moderate models, \textit{I\_B\_timing\_range} contributed more explanatory power than the other factors. For \textit{P\_B\_usage\_group}, we observe a positive relation with the issue-addressing likelihood in Figure~\ref{fig:rq3_trend_two}. One possible explanation is that projects with a higher usage of bounties, have more experience in using bounties which may increase the likelihood of a bounty being addressed. For example, the \code{eslint} project maintains a document on how bounties work\footnote{\url{http://bit.ly/2Sql57n}}. The \code{eslint} project has 43 successful (i.e., closed-paid) and only one failed (i.e., open-unpaid) bounty issue report.

\textbf{The total bounty value of an issue report is the most important factor that impacts the issue-addressing likelihood in the first-timer bounty-projects, while it is less important in the projects in which bounties are used more frequently.} From Table~\ref{tab:top3impFactors}, we can see that \textit{I\_B\_total\_value} (i.e., the total bounty value of a bounty issue report) is the most important factor in the first-time model, while it is not as important in projects in which bounties are used more frequently. Figure~\ref{fig:rq3_trend_two} shows a positive relation between \textit{I\_B\_total\_value} and the issue-addressing likelihood of the first-time projects. This observation is compatible with our observations in Section~\ref{rq1} that successful bounty issue reports usually have much higher bounty values than failed bounty issue reports in the first-timer bounty-projects.

\rqbox{
In general, the timing of proposing bounties and the bounty-usage frequency are the top two most important factors that impact the issue-addressing likelihood. The total bounty value that an issue report has is the most important factor that impacts the issue-addressing likelihood in the first-timer bounty-projects, while it is not as important for projects in which bounties are frequently used.
}

\section{Discussion}
\label{sec:dis}
\subsection{Studying ignored bounty issue reports}
\label{rq4}

In Section~\ref{prestudy}, we observed that in 19.7\% of the bounty issue reports the bounties were ignored (i.e., closed-unpaid). For these ignored bounty issue reports, the issue reports were closed but the bounties remained unclaimed.
It seems that money was not the driver that motivated developers to address these issues.
To understand the possible reasons behind this phenomenon, we manually studied all 692 ignored bounty issue reports (with a total bounty value of \$41,856).
Because the ``closed'' status of an issue report does not necessarily mean that the issue was addressed (e.g., a report may have been a duplicate of another issue report), it is very difficult to automatically identify whether an issue in the closed issue report was addressed or not.
Therefore, we need to manually examine the closed-unpaid bounty issues reports to filter out the reports that were closed for another reason than the issue being addressed.

\textbf{21.8\% (479 out of 2,200) of the addressed bounty issue reports were not paid out.} We identified that 479 out of the studied 692 bounty issue reports were closed because the issues were addressed. Such cases are interesting since the developers could have claimed the bounty but they did not. We manually examined the discussion for these 479 issue reports. We identified 19 cases in which developers gave an explanation for not claiming the bounty. We grouped the explanations as follows:

\noindent\textbf{The developer is not driven by money.} In 7 out of 19 cases a developer refused to claim the bounty because they were not motivated by money to address the issue.
For example, one developer was against the bounty because they felt that the issue-addressing process should be driven by the interests of the community rather than money. A contributor of the \code{Brython} project, refused the bounty because he wanted to keep \code{Brython} free from monetary motivations: ``\textit{What is this `bounty' thing? Needless to say, I refuse that anybody (me included, of course) gets paid for anything related to \code{Brython}.}''\footnote{\url{http://bit.ly/2OTYx0x}} In addition, he also asked bounty backers to remove all bounties within the \code{Brython} project although he respected prior bounties that were paid out. There were five bounty issue reports in the \code{Brython} project and four bounty issue reports that were addressed without claiming the bounty.

\noindent\textbf{The developer is afraid of sending the wrong message.} Krishnamurthy et al. \cite{krishnamurthy2006bounty} pointed out that financial incentives may cause confusion in the community because the financial incentives may drive a project's own product development cycle away from what is in place.
We observed that developers expressed similar concerns. A developer of the \code{Facebook/HHVM} project, explained that: ``\emph{That's very generous of you, but I can't accept a bounty for doing my job. :-P It would be a conflict of interest, and I worry it sends the wrong message about how we prioritize issues from the community.}''\footnote{\url{http://bit.ly/2OZw1uw}}

\noindent\textbf{The issue report was addressed by more than one developer.} We found nine cases where bounties ended up unclaimed because an issue report was addressed by multiple developers cooperatively and they felt inappropriate to claim the bounty by one developer. For example, the issue\footnote{\url{http://bit.ly/2PrMiHV}} was addressed by two developers and because a bounty cannot be split into two parts, no one claimed it.

\subsection{The implications of our findings}
\textbf{Backers should consider proposing a bounty as early as possible and be cautious when proposing small bounties on long-standing issue reports.}
The timing of proposing a bounty is an important factor that impacts the issue-addressing likelihood. In Sections~\ref{rq2} and~\ref{rq3}, we revealed the fact that issue reports for which bounties were proposed earlier are more likely to be addressed. Additionally, we observed that issue reports for which bounties were proposed earlier are more likely to be addressed faster. Backers benefit from the higher issue-addressing likelihood and faster issue-addressing speed by proposing bounties earlier.

In Section~\ref{rq2}, we also noticed a big drop (i.e., from 53.2\% to 30.1\%) of the issue-addressing likelihood when backers proposed bounties on long-standing (i.e., more than half a year) issue reports. This drop might be due to such issue reports having become obsolete or being hard to address.
Since bounties with a value of less than \$100 will not be refunded to the backers if the issue report remains unaddressed, we suggest that backers be cautious when proposing small bounties on long-standing issue reports.

\textbf{Backers should consider proposing a relatively bigger bounty in first-timer bounty-projects.} Although the issue-addressing likelihood is only 37.4\%  for projects with no bounty-usage experience, the first-timer model in Section~\ref{rq3} shows that the bounty value of an issue report is the most important factor in the first-timer projects, as the issue-addressing likelihood is higher for higher bounty values. The high ratio (2.5) of the bounty value of successful bounty issue reports to the bounty value of failed bounty issue reports also supports this finding.
We suggest that backers of projects with no bounty-usage experience propose higher bounty values for issue reports.

\textbf{Bounty platforms should allow for splittable multi-hunter bounties.}
In addition to a voluntary nature, open source projects have a collaborative nature. Some issues are hard for a developer to address alone. Hence, we encourage developers to work together, especially for issue reports which have a high bounty value (as these issue reports are often harder to address).
However, the current bounty workflow only allows \textbf{one} bounty hunter to claim the bounty, which goes against the collaborative nature of open source. It may also drive the developers, who want to collaboratively address the issue, away because not every participant will get a reward at the end. Therefore, bounty platforms should consider adding the ability for a bounty to be split across multiple hunters to encourage developers to work together on difficult bounty issues.

\textbf{Bounties should be transferable.}
The total value of all addressed-unpaid bounties (\$43,256) is ``frozen'' in Bountysource. In addition, the median number of days between the closing date of the issue report and the date of collecting our data is 372.5,  which means that more than half of the bounties from the ignored bounty issue reports were unclaimed for at least one year.
By manually examining these 479 addressed-unpaid bounty issue reports, we found 31 cases in which someone reminded the bounty hunter to claim the bounty, however, the reminder was ignored. By reassigning these unclaimed bounties to other issue reports, a larger value could be created for these ``stale'' bounties. For example, Bountysource can suggest and enable backers to assign their long-standing unclaimed bounties to another unaddressed issue report, which has many comments (i.e., people care about it), to encourage developers to address the issue report. Interestingly, we also found suggestions from developers who did not want to receive the bounty but suggested the bounty backers to transfer the bounty to other issue reports or to the project as a kind of funding.

\section{Threats to Validity}
\label{sec:threats}
In this section, we discuss the threats to validity.
Threats to \textbf{external validity} are related to the generalizability of our findings. We studied only bounty issue reports from GitHub and Bountysource. Further research is necessary to find whether our findings are generalizable to other types of issue reports (e.g., from commercial platforms) and other bounty platforms.
Although our models have a high explanatory power, there might be additional factors that relate to the likelihood of an issue being addressed. Future studies should investigate more factors.

Threats to \textbf{internal validity} relate to the experimenter bias and errors.
One threat relates to the project categorization in Section~\ref{rq1}, in which we used 50 bounty issue reports as a threshold to distinguish whether a project uses bounties moderately or frequently. To alleviate this threat, we redid the analysis of Section~\ref{rq1} with other bounty-usage frequency thresholds (i.e., 40 and 60). The resuls show that our findings still hold (see our online appendix~\cite{appendix} for more details).

Another threat is that we rely on manual analysis to identify the addressed-unpaid issues and to identify why developers did not claim a bounty in Section~\ref{rq4}, which may introduce bias due to human factors. To mitigate the threat of bias during the manual analysis, two of the authors conducted the manual analysis and discussed conflicts until a consensus was reached. We used Cohen's kappa~\cite{cohenkappa} to measure the inter-rater agreement and the value is 0.86, which indicates a high level of agreement.

\section{Related work}
\label{relatedwork}
In this section, we discuss related work along two dimensions: the bounty in software engineering and the improvement of the issue-addressing process.

\textit{Bounties in software development:}
Bounties are used to attract developers and motivate them to complete tasks. Prior work has studied the impact of bounties on software development. \cite{krishnamurthy2006bounty} gave an overview of bounties in Free/Libre/Open Source Software (FLOSS).
They observed that bounty hunters' responses are related to the workload, the probability of winning the bounty, the value of the bounty and the recognition that they might receive by winning the bounty. Different from their study, we focused on using bounties to improve the issue addressing process.

Several studies focused on the usage of bounties to motivate developers to detect software security vulnerabilities.
Finifter et al.~\cite{finifter2013empirical} analyzed vulnerability rewards programs for Chrome and Firefox. They found that the rewards programs for both projects are economically effective, compared to the cost of hiring full-time security researchers.
Zhao et al.~\cite{zhao2014exploratory} investigated the characteristics of hunters in bug-bounty programs and found that the diversity of hunters improved the productivity of the vulnerability discovery process. Hata et al.~\cite{hata2017understanding} found that most hunters are not very active (i.e., they have only a few contributions). These findings are similar to our finding that most hunters only addressed one bounty issue. Zhao et al. and Maillart et al.~\cite{zhao2017devising,maillart2017given} analyzed the effect of different policies of bug-bounty programs.
By studying bug-bounties from several perspectives, they provided insights on how to improve the bug-bounty programs. For example, Maillart et al.~\cite{maillart2017given} suggested project managers to dynamically adjust the value of rewards according to the market situation (e.g., increase rewards when releasing a new version).

However, there is not much research to study the effectiveness of bounties in the issue-addressing process.
The work of Kanda et al.~\cite{kanda2017towards} is closest to ours. They studied GitHub and Bountysource data, but studied only 31 projects (compared to 1,203 in our study). They compared the closed-rate and closing-time between bounty issue and non-bounty issue reports. Their results showed that the closing-rate of bounty issue reports is lower than that of non-bounty issue reports, and it takes longer for the bounty issue reports to get closed than non-bounty issue reports.
Our study performs a deeper analysis of bounties at the project and the time level. Besides, we further study the relationship between the issue-addressing likelihood and the bounty-related factors (e.g., the total bounty value of a bounty issue report) while controlling for the factors that are related to the issue report and project (e.g., the number of comments before the first bounty is proposed). We found that the timing and the bounty-usage frequency are the most important factors in increasing the issue-addressing likelihood.

\textit{Improving the issue-addressing process:}
Issue addressing is an essential activity in the life cycle of software development and maintenance. Therefore, a large amount of research was done to improve the issue-addressing process.
One group of studies focused on providing insights into improving the issue-addressing process in aspects of the quality of issue reports, the effectiveness of developers and automated bug localization and fixing.
For example, Bettenburg et al.~\cite{bettenburg2008makes,hooimeijer2007modeling} analyzed the quality of bug reports (i.e., a type of issue report) and provided some guidelines for users to generate high-quality reports so that developers can address issues more efficiently. Ortu et al.~\cite{Ortu:2015} analyzed the relation between sentiment, emotions, and politeness of developers in comments with needed time to address an issue. They found that the happier developers are, the shorter the issue-addressing time is likely to be. Zhong et al.~\cite{Zhong:2015} performed an empirical study on real-world bug fixes to provide insights and guideline for improving the state-of-the-art of automated program repair. Soto et al.~\cite{Soto:2016} performed a large-scale study of bug-fixing commits in Java projects and provided insights for high-quality automated software repair to target Java code.
A number of studies helped developers locate the buggy code in projects using information retrieval techniques~\cite{Zhou:2012,Saha:2013,WangL16,WangLL14}.
Different from prior studies, we perform an empirical study to understand the relationship between bounties and the issue-addressing process. We provide insights into how to better use the bounty to improve the issue-addressing process.

\section{Conclusion}
\label{conclusion}

In this paper, we studied 5,445 bounties with a total value of \$406,425 from Bountysource along with their associated 3,509 issue reports from GitHub to study the relationship between the bounty (e.g., timing of proposing a bounty, bounty value, and bounty-usage frequency) and the issue-addressing likelihood.
We found that: 1) The timing of proposing bounties and the bounty-usage frequency are the most important factors that impact the issue-addressing likelihood. Issue reports for which bounties were proposed earlier are more likely to have a higher issue-addressing likelihood and a faster addressing-speed. 2) In first-timer bounty-projects, the issue-addressing likelihood is higher for higher bounty values and in these projects, backers should consider proposing a relatively bigger bounty. 3) Backers should be cautious when proposing small bounties on long-standing issue reports as they risk losing money without getting their issue addressed.

Our findings suggest that backers should consider proposing a bounty early and be cautious when proposing small bounties on long-standing issue reports. Bounty platforms should allow dividing bounties between hunters, and transferring bounties to other issue reports.

\balance
\bibliographystyle{abbrv}
{
\footnotesize
\bibliography{ref}

\begin{thebibliography}{10}

\bibitem{bauer1972constructing}
D.~F. Bauer.
\newblock Constructing confidence sets using rank statistics.
\newblock {\em Journal of the American Statistical Association},
  67(339):687--690, 1972.

\bibitem{bettenburg2008makes}
N.~Bettenburg, S.~Just, A.~Schr{\"o}ter, C.~Weiss, R.~Premraj, and
  T.~Zimmermann.
\newblock What makes a good bug report?
\newblock In {\em Proc. of the Int'l Symp. on Foundations of Software
  Engineering}, pages 308--318. ACM, 2008.

\bibitem{efron1986biased}
B.~Efron.
\newblock How biased is the apparent error rate of a prediction rule?
\newblock {\em Journal of the American Statistical Association},
  81(394):461--470, 1986.

\bibitem{finifter2013empirical}
M.~Finifter, D.~Akhawe, and D.~Wagner.
\newblock An empirical study of vulnerability rewards programs.
\newblock In {\em USENIX Security Symp.}, pages 273--288, 2013.

\bibitem{cohenkappa}
K.~Gwet et~al.
\newblock Inter-rater reliability: dependency on trait prevalence and marginal
  homogeneity.
\newblock {\em Statistical Methods for Inter-Rater Reliability Assessment
  Series}, 2:1--9, 2002.

\bibitem{Harrell:2006}
F.~E. Harrell, Jr.
\newblock {\em Regression Modeling Strategies}.
\newblock Springer-Verlag New York, Inc., Secaucus, NJ, USA, 2006.

\bibitem{hata2017understanding}
H.~Hata, M.~Guo, and M.~A. Babar.
\newblock Understanding the heterogeneity of contributors in bug bounty
  programs.
\newblock In {\em Proc. of the ACM/IEEE Int'l Symp. on Empirical Software
  Engineering and Measurement}, pages 223--228, 2017.

\bibitem{hooimeijer2007modeling}
P.~Hooimeijer and W.~Weimer.
\newblock Modeling bug report quality.
\newblock In {\em Proc. of the Int'l Conf. on Automated Software Engineering},
  pages 34--43. ACM, 2007.

\bibitem{kabinna2018examining}
S.~Kabinna, C.-P. Bezemer, W.~Shang, M.~D. Syer, and A.~E. Hassan.
\newblock Examining the stability of logging statements.
\newblock {\em Empirical Software Engineering}, 23(1):290--333, 2018.

\bibitem{kanda2017towards}
T.~Kanda, M.~Guo, H.~Hata, and K.~Matsumoto.
\newblock Towards understanding an open-source bounty: Analysis of
  {B}ountysource.
\newblock In {\em Int'l Conf. on Software Analysis, Evolution and
  Reengineering}, pages 577--578. IEEE, 2017.

\bibitem{krishnamurthy2006bounty}
S.~Krishnamurthy and A.~K. Tripathi.
\newblock Bounty programs in free/libre/open source software.
\newblock In {\em The Economics of Open Source Software Development}, pages
  165--183. Elsevier, 2006.

\bibitem{long2003ordinal}
J.~D. Long, D.~Feng, and N.~Cliff.
\newblock Ordinal analysis of behavioral data.
\newblock {\em Handbook of psychology}, 2003.

\bibitem{maillart2017given}
T.~Maillart, M.~Zhao, J.~Grossklags, and J.~Chuang.
\newblock Given enough eyeballs, all bugs are shallow? {Revisiting Eric Raymond
  with bug bounty programs}.
\newblock {\em Journal of Cybersecurity}, 3(2):81--90, 2017.

\bibitem{mcintosh2016empirical}
S.~McIntosh, Y.~Kamei, B.~Adams, and A.~E. Hassan.
\newblock An empirical study of the impact of modern code review practices on
  software quality.
\newblock {\em Empirical Software Engineering}, 21(5):2146--2189, 2016.

\bibitem{Ortu:2015}
M.~Ortu, B.~Adams, G.~Destefanis, P.~Tourani, M.~Marchesi, and R.~Tonelli.
\newblock Are bullies more productive?: Empirical study of affectiveness vs.
  issue fixing time.
\newblock In {\em Proc. of the Working Conf. on Mining Software Repositories},
  pages 303--313, 2015.

\bibitem{gopi:2017}
G.~K. Rajbahadur, S.~Wang, Y.~Kamei, and A.~E. Hassan.
\newblock The impact of using regression models to build defect classifiers.
\newblock In {\em Proc. of the Int'l Conf. on Mining Software Repositories},
  pages 135--145, 2017.

\bibitem{romano2006appropriate}
J.~Romano, J.~D. Kromrey, J.~Coraggio, and J.~Skowronek.
\newblock Appropriate statistics for ordinal level data: Should we really be
  using t-test and {Cohen's} d for evaluating group differences on the {NSSE}
  and other surveys.
\newblock In {\em Annual meeting of the Florida Association of Institutional
  Research}, pages 1--33, 2006.

\bibitem{Saha:2013}
R.~K. Saha, M.~Lease, S.~Khurshid, and D.~E. Perry.
\newblock Improving bug localization using structured information retrieval.
\newblock In {\em Proc. of the IEEE/ACM Int'l Conf. on Automated Software
  Engineering}, pages 345--355, 2013.

\bibitem{Soto:2016}
M.~Soto, F.~Thung, C.-P. Wong, C.~Le~Goues, and D.~Lo.
\newblock A deeper look into bug fixes: Patterns, replacements, deletions, and
  additions.
\newblock In {\em Proc. of the Int'l Conf. on Mining Software Repositories},
  pages 512--515, 2016.

\bibitem{wang2017understanding}
S.~Wang, T.-H. Chen, and A.~E. Hassan.
\newblock Understanding the factors for fast answers in technical {Q\&A}
  websites.
\newblock {\em Empirical Software Engineering}, pages 1--42, 2017.

\bibitem{WangL16}
S.~Wang and D.~Lo.
\newblock Amalgam+: Composing rich information sources for accurate bug
  localization.
\newblock {\em Journal of Software: Evolution and Process}, 28(10):921--942,
  2016.

\bibitem{WangLL14}
S.~Wang, D.~Lo, and J.~Lawall.
\newblock Compositional vector space models for improved bug localization.
\newblock In {\em {IEEE} Int'l Conf. on Software Maintenance and Evolution},
  pages 171--180, 2014.

\bibitem{zhao2014exploratory}
M.~Zhao, J.~Grossklags, and K.~Chen.
\newblock An exploratory study of white hat behaviors in a web vulnerability
  disclosure program.
\newblock In {\em Proc. of the Workshop on Security Information Workers}, pages
  51--58. ACM, 2014.

\bibitem{zhao2017devising}
M.~Zhao, A.~Laszka, and J.~Grossklags.
\newblock Devising effective policies for bug-bounty platforms and security
  vulnerability discovery.
\newblock {\em Journal of Information Policy}, 7:372--418, 2017.

\bibitem{Zhong:2015}
H.~Zhong and Z.~Su.
\newblock An empirical study on real bug fixes.
\newblock In {\em Proc. of the Int'l Conf. on Software Engineering}, pages
  913--923, 2015.

\bibitem{appendix}
J.~Zhou.
\newblock Supplementary material for our paper.
\newblock \url{https://github.com/yiikou/Bountysource/appendix.pdf}, 2018.

\bibitem{Zhou:2012}
J.~Zhou, H.~Zhang, and D.~Lo.
\newblock Where should the bugs be fixed? - more accurate information
  retrieval-based bug localization based on bug reports.
\newblock In {\em Proceedings of the 34th International Conference on Software
  Engineering}, pages 14--24, 2012.

\end{thebibliography}
}

\end{document}